\documentclass[10pt,letterpaper]{article}
\usepackage{opex3}

\begin{document}

%%%%%%%%%%%%%%%%%% title page information %%%%%%%%%%%%%%%%%%
\title{Spatial properties of twin-beam correlations at low- to high-intensity transition}

\author{Radek Machulka$^{1}$, Ond\v{r}ej Haderka$^{2,*}$, Jan Pe\v{r}ina Jr.$^{1}$,\\
Marco Lamperti,$^{3}$ Alessia Allevi,$^{3,4}$ and Maria Bondani$^{5,4}$}

\address{
$^1$ RCPTM, Joint Laboratory of Optics of PU and Inst. Phys. AS
CR, 17. listopadu 12, 77146 Olomouc, Czech Republic \\
$^2$ Institute of Physics AS CR, Joint Laboratory of Optics,
17. listopadu 50a, 77146 Olomouc, Czech Republic \\
$^3$Dipartimento di Scienza e Alta Tecnologia, Universit\`a degli Studi
dell'Insubria, Via Valleggio 11, 22100 Como, Italy\\
$^4$CNISM UdR Como, via Valleggio 11, 22100 Como, Italy\\
$^5$Istituto di Fotonica e Nanotecnologie, CNR, Via Valleggio 11, 22100 Como, Italy
}
\email{$^*$ondrej.haderka@upol.cz} %% email address is required



\begin{abstract}
It is shown that spatial correlation functions measured for
correlated photon pairs at the single-photon level correspond to
speckle patterns visible at high intensities. This correspondence
is observed for the first time in one experimental setup by using
different acquisition modes of an intensified CCD camera in low
and high intensity regimes. The behavior of intensity auto- and
cross-correlation functions in dependence on pump-beam parameters
including power and transverse profile is investigated.
\end{abstract}


%%%%%%%%%%%%%%%%%%%%%%% References %%%%%%%%%%%%%%%%%%%%%%%%%

%%%%%%%%%%%%%%%%%%%%%%%%%%  body  %%%%%%%%%%%%%%%%%%%%%%%%%%
\section{Introduction}
Optical twin-beams generated by spontaneous parametric
down-conversion (PDC) are well known to exhibit temporal and
spatial correlations \cite{Mandel}. The first experimental
measurements of spatial correlations in twin beams, aimed at
determining the size of the coherence areas in the transverse
planes, date back to 1990s \cite{Malygin1985,Jost1998} and early
2000s \cite{Lantz2004,Haderka2005}. Such measurements were
performed in the single-photon regime on twin-beams containing
one photon at most by using scanned single-photon detectors
(photomultipliers operating in Geiger mode \cite{Malygin1985},
avalanche photodiodes \cite{Monken1998,Molina2005}), intensified
CCDs \cite{Jost1998,Haderka2005} or EMCCDs \cite{Edgar2012}). The experimental results can be
compared to a well established theory properly working in this
regime [9-12].
%% \cite{Grayson1994,Joobeur1994,Joobeur1996,Hamar2010}.

On the other hand, twin-beam states can also be generated in a much
higher intensity regime \cite{Brambilla2004}, in which coherence
areas become visible in single-shot images \cite{BondaniEPJST,Jedrkiewicz04}.
In the high-gain parametric process, the sizes of the coherence areas exhibit
dependence on the gain \cite{Jedrkiewicz04}.
The quantification of the size of coherence areas can be obtained
through intensity auto- (AC, $ \Gamma_{ss} $,  $ \Gamma_{ii} $)
and cross-correlation (XC, $ \Gamma_{si} $,  $ \Gamma_{is} $)
functions \cite{Brida2009a,Brida2009b} defined as
\begin{equation}  % 1
 \Gamma_{ab}({\bf r}_1,{\bf r}_2) = \langle \Delta I_a({\bf r}_1)
  \Delta I_b({\bf r}_2) \rangle , \hspace{2mm} a,b = s,i .
\label{1}
\end{equation}
Symbol $ s $ [$ i $] stands for a signal [idler] beam with
intensity fluctuation $ \Delta I_s({\bf r}_1) $ [$ \Delta I_i({\bf
r}_2) $] at position $ {\bf r}_1 $ [$ {\bf r}_2 $] and $ \langle
\rangle $ denotes quantum (statistical) averaging. Although the
two described intensity regimes together with their own
experimental techniques differ considerably, the observed effects
arise from the same nonlinear process. That is why it is important
to compare quantitatively the observed coherence properties at
single-photon and intense twin-beam regime.

In a recent publication \cite{Hamar2010} we explored in detail
spatial coherence of twin beams at single-photon level by using
photon-counting methods. As the detector we used a single-photon
sensitive intensified CCD camera illuminated by a PDC field
containing much less than one photon per active superpixel and
detection interval of the camera. In particular, we have shown
that the transverse coherence functions of twin beams can be
tailored by changing the pump-beam properties.

%%%%%%%%%
\begin{figure}[tbh]   % 1
\centering{\includegraphics[width=6cm]{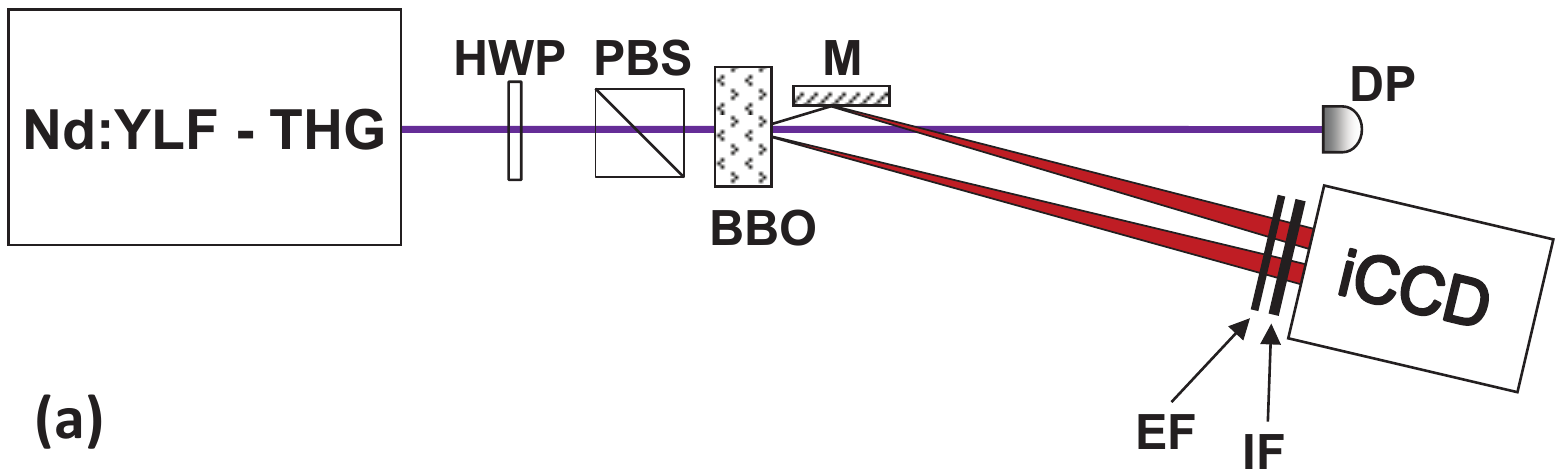}
\hspace{2mm}
\includegraphics[width=6cm]{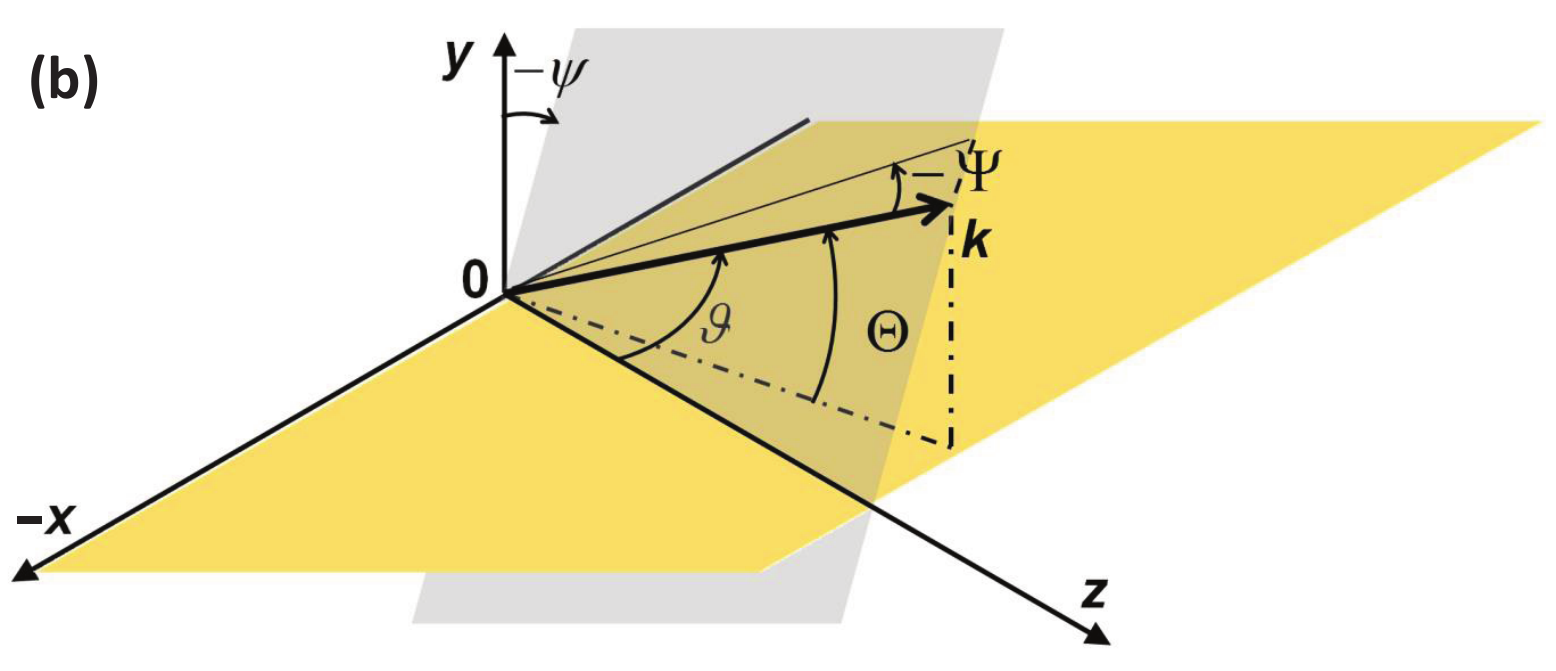}}
\caption{(a) Sketch of the experimental setup (see text for details).
(b) Propagation geometry of a plane wave with wave vector $ {\bf k} $ in the
reference frame of the crystal: $ \vartheta $ is the radial angle measured
from the $ z $ axis in the incidence plane and $ \psi $ is the azimuthal angle that gives
the rotation of the plane of incidence with respect to the plane $ yz $.
Alternatively, the coordinate system of the camera is used: $ \Theta $ is
the radial angle measured in the plane perpendicular to the $ xz $ plane
from the intersection of the two planes and $ \Psi $ is the azimuthal angle
in the plane perpendicular to the plane of angle $ \Theta $.}
\label{setup}
\end{figure}
%%%%%%%%%

In this paper, we investigate the above mentioned regime hand in
hand with the intense twin-beam regime containing formed speckles.
In the latter regime, the PDC field is so intense that many
photons impinge on each superpixel of the camera during the
detection interval. As a consequence, the camera amplification is tuned
down and the camera loses its
single-photon sensitivity and provides only classical field
intensities. Nevertheless, the intensity XC
functions evaluated from intensity measurements correspond to
those measured by the photon-counting technique at the single-photon
level. Moreover, the dependence of both intensity AC and XC
functions on pump-beam parameters fits with
that observed at single-photon level and predicted by theory.

\section{Experimental setup}
Our experiment was performed by using the third-harmonics
(349 nm) of a ps-pulsed mode-locked Nd:YLF laser (High-Q Laser)
to pump a 8-mm-long BBO crystal ($\theta=37^\circ$). Nearly
degenerate photon pairs occurred at a cone layer with a vertex
half-angle of $11.9^\circ$. A part of the cone layer was selected by
a 40-nm-wide bandpass interference filter (IF)
centered at 710~nm, that is close to frequency degeneracy. One part
of the cone-layer (signal) was captured directly by the photocathode
of the iCCD camera (Andor DH734, pixel size $13\times13~\mu$m$^2$)
placed 38.5~cm behind the crystal, while the corresponding twin
portion (idler) was directed to a different section of the photocathode using a 
dielectric mirror (M, see Fig.~\ref{setup}). An additional long-pass filter (EF) was
used to cut scattered pumping photons.
The camera was triggered by an electronic pulse derived from the laser
pulse and set to 5~ns detection window (used both for low- and high-intensity
measurement) to assure detection of PDC from a single pump pulse and to minimize noise.
The power of the transmitted pump beam was
monitored using a power-meter. Moreover, the pump beam diameter
was measured in the horizontal and vertical directions using a
scanning knife-edge. The pump beam was nearly elliptical with the
vertical width reaching approximately 70\% of the horizontal one.
The pump power was changed by using a half-wave plate (HWP) followed
by a polarizing beam splitter.

%%%%%%%%%
\begin{figure}[tbh]   % 2
\centering{(a)\hspace{2mm}\includegraphics[width=3.3cm,height=3.3cm,clip]{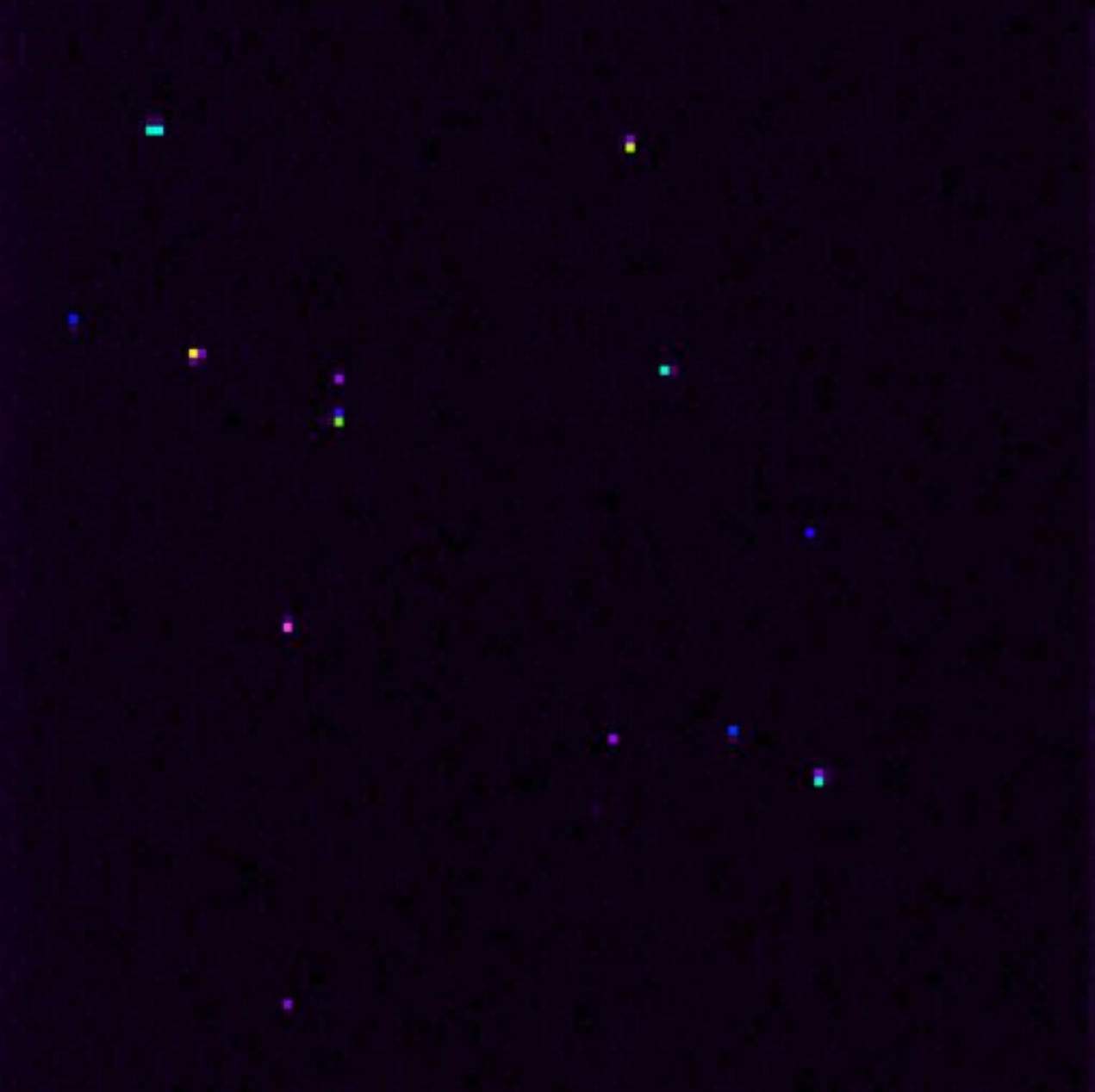}
\hspace{2mm}
(b)\hspace{2mm}\includegraphics[width=3.3cm,height=3.3cm,clip]{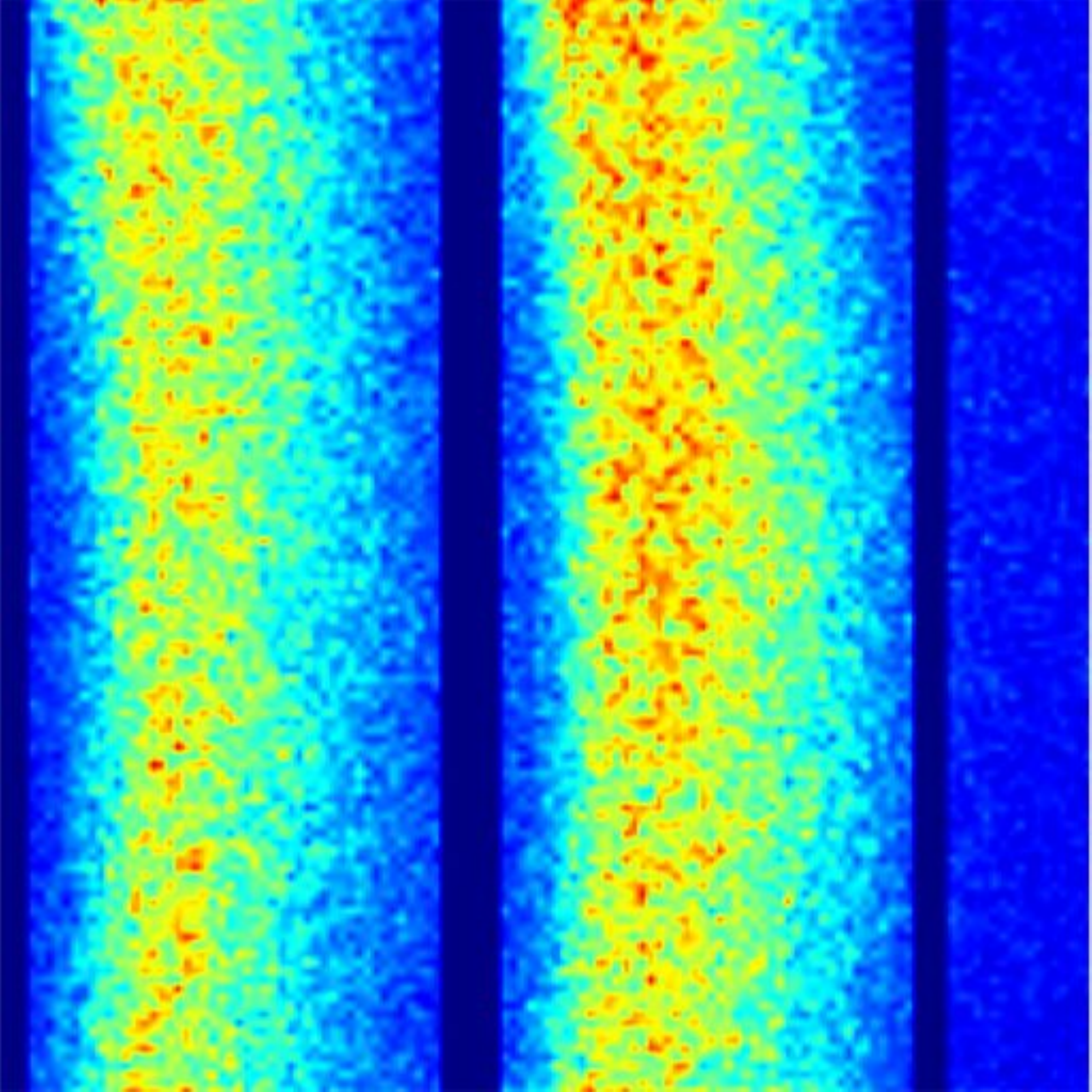}}  \\

\centering{(c)\hspace{2mm}\includegraphics[width=3.3cm,height=3.3cm,clip]{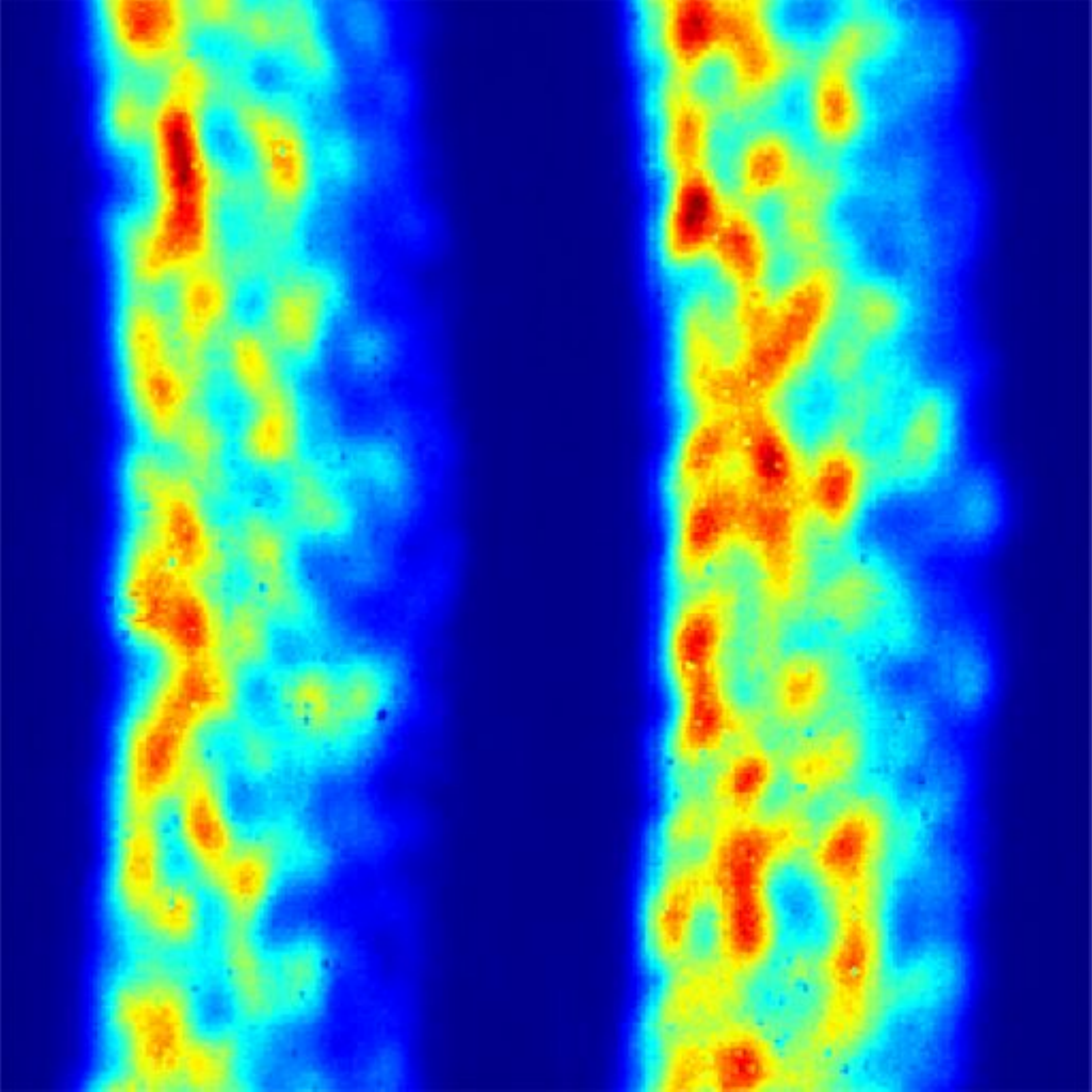}
\hspace{2mm}
(d)\hspace{2mm}\includegraphics[width=3.3cm,height=3.3cm,clip]{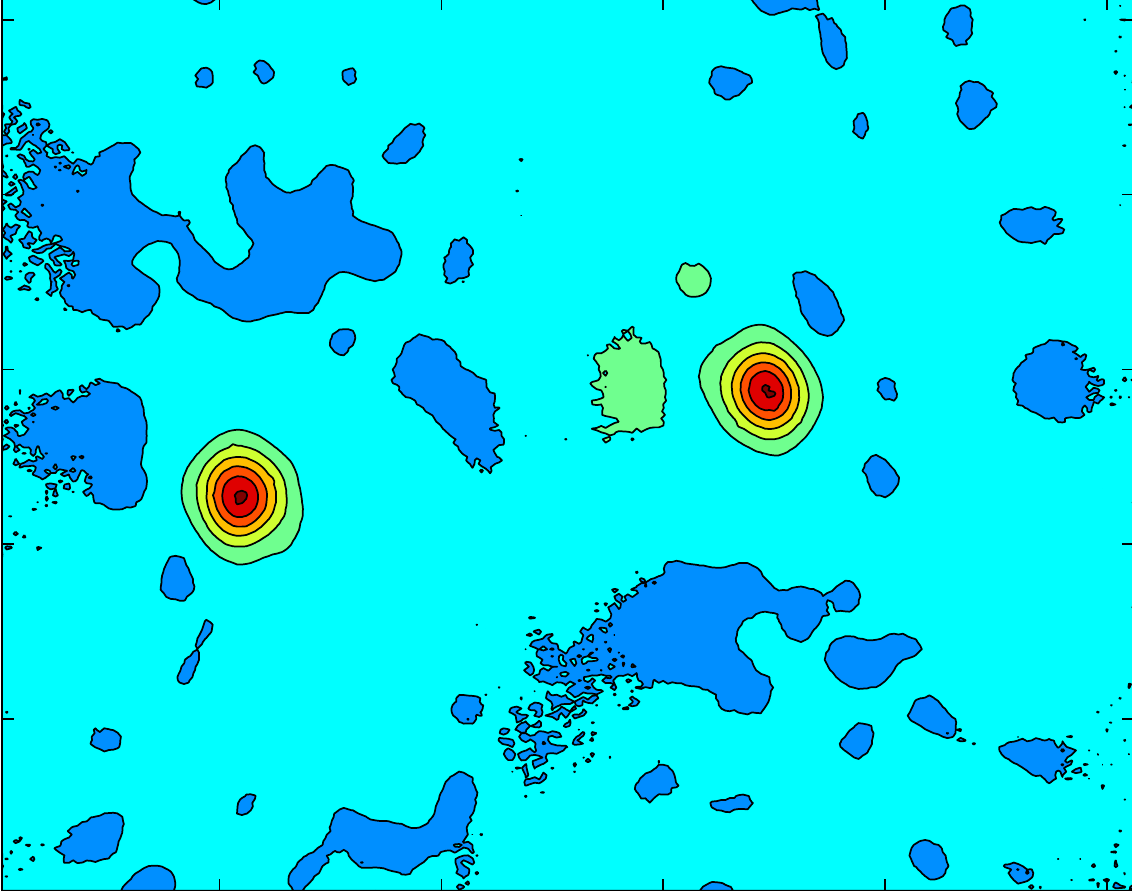}}
\caption{Typical images taken by the iCCD camera in the
experiment. a) single frame at the single-photon level
($P_{p}=20~\mu$W). b) summed image of frames at the single-photon
level with the overall mean photon number equal to that in c); c) single frame
at the mesoscopic level with speckles ($P_{p}=45$~mW). d)
intensity AC and XC functions computed from a sequence of 1000
frames taken at classical intensity. Two
peaks appear in the image: the left peak is located around
the selected idler point, thus describing the intensity AC function
$\Gamma_{ii}$. The right peak is centered at the conjugate signal
point given by ideal phase-matching conditions, thus giving the
intensity XC function $\Gamma_{is}$. The axes are scaled in
superpixels of the iCCD.} \label{images}
\end{figure}
%%%%%%%%%

\section{Results}
First of all, we measured the correlation area characterizing
intensity XC function in the transverse plane by exploiting the
photon-counting method presented in \cite{Hamar2010}. We took
a long sequence of camera frames [see Figs.~2(a) and 2(b)]
and processed them in the way described in \cite{Haderka2005}.
The camera was set to its maximum gain and a 8x8-hardware
binning was used to speed up the acquisition. In each frame of the
sequence, single-photon detection events have been identified separately
for the signal and idler regions after thresholding above the readout noise
of the CCD chip. Since the number of binned superpixels ($>6000$) is much larger than
the number of incident photons, the impinging probability is well below
one photon per superpixel and the camera serves in fact as two (signal and idler)
photon-number-sensitive detectors in this regime. Using the whole sequence,
correlations of detection-event positions are obtained as well as first
and second moments of the photocount distributions. From these data we can get the 
correlation area with a full-width half maximum (FWHM)
of $\Delta\Gamma_{is,\Theta}=(490\pm52)~\mu$m in the horizontal
direction (radial with respect to the cone section) and
$\Delta\Gamma_{is,\Psi}=(710\pm52)~\mu$m in the vertical (azimuthal)
one. The sequence was measured at the pump power $ P_p =
20~\mu$W, which resulted in 8.9 mean idler detected photons
and 10.5 mean signal detected photons. As our method naturally
includes also the determination of effective quantum detection
efficiency \cite{Perina2012}, we could assign 7.2\% and 8.5\% detection
efficiency to idler and signal beams,
respectively. We ascribe this difference to the non-perfect
reflectivity of the mirror in the idler path.

The increase of pump power $ P_{p} $ causes the transition from
photon-counting to classical-intensity measurement and patterns
with speckles [see Fig.~2(c)] occur in single-shot images for
$ P_p $ values larger than 20~mW. By setting a 4x4-hardware binning, 
we took a sequence of 1000 images for
different pump powers and pump-beam sizes and processed them by
computing intensity AC functions $\Gamma_{ii}$ and XC functions
$\Gamma_{is}$ at 100 randomly selected points in the fringe
patterns close to the frequency degeneracy of the PDC process
[for an example, see Fig.~2(d)]. This approach gives a better signal-to-noise
ratio than taking the correlation of the whole image, as only a part of the
image is covered by a significant light intensity. These correlation functions are
conveniently characterized by their horizontal
$\Delta\Gamma_{\Theta}$ and vertical $\Delta\Gamma_{\Psi}$ widths
that depend on pump-beam parameters.

%%%%%%%%%
\begin{figure}[tbh] % 3
\centering{(a)\includegraphics[width=5.5cm]{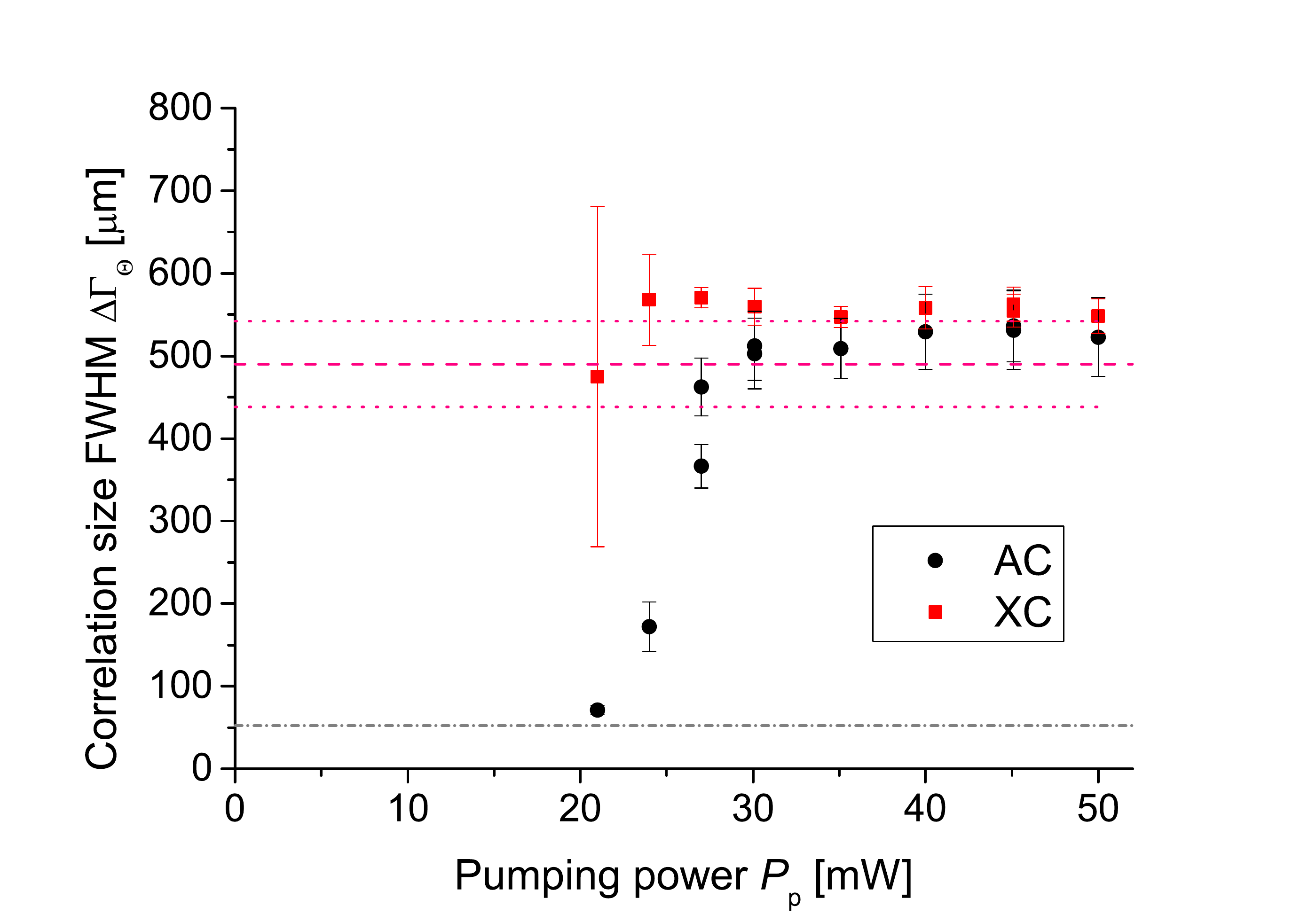}
(b)\includegraphics[width=5.5cm]{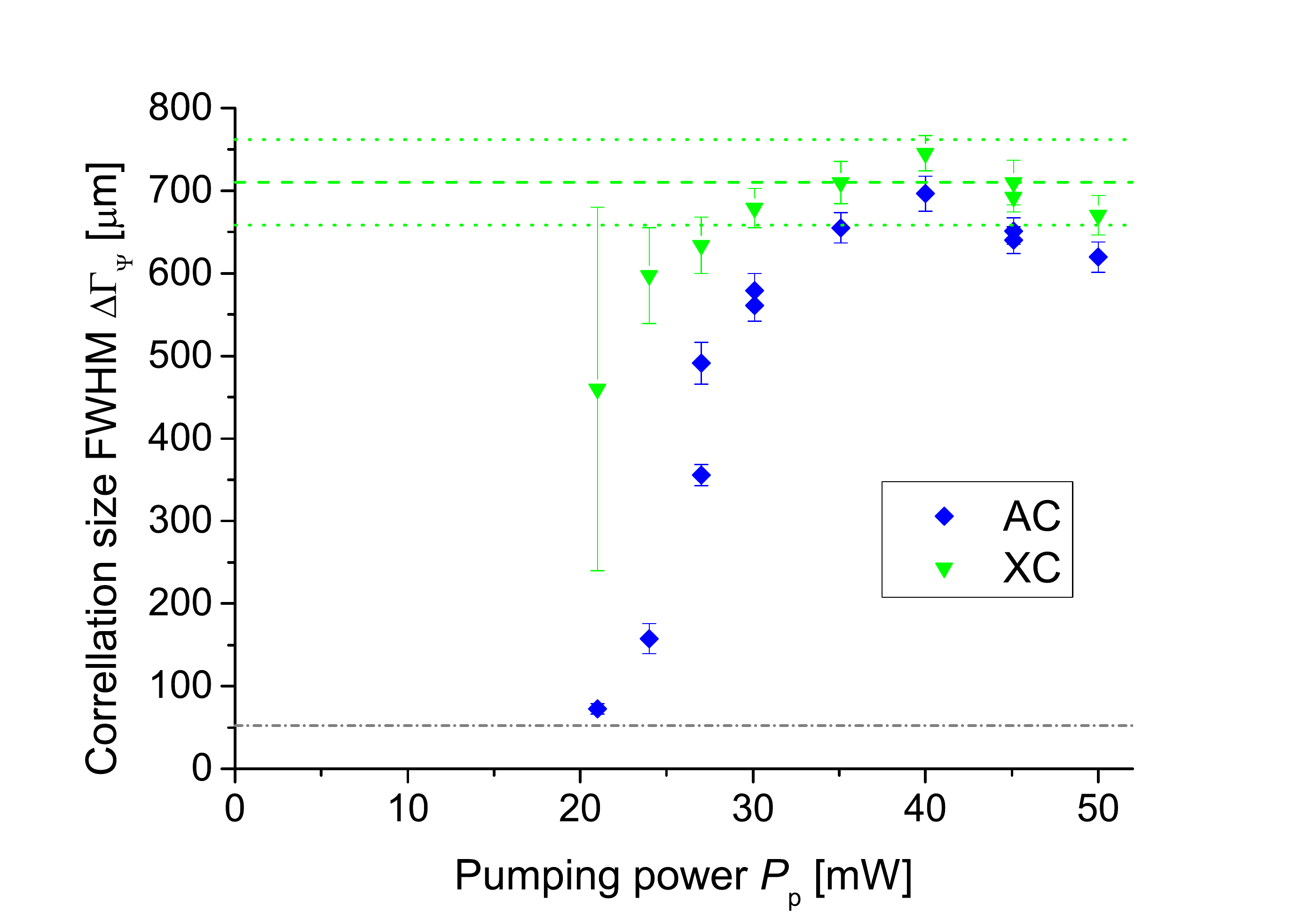}}
\caption{(a) Radial width $\Delta\Gamma_{\Theta}$ and (b)
azimuthal width $\Delta\Gamma_{\Psi}$ of intensity AC (rectangles)
and XC (circles) functions (FWHM) depending on pump power $ P_{p}
$. Dashed lines correspond to the XC function width $\Delta\Gamma
$ measured at power $ P_{p} = 20~\mu$W by the photon-counting
approach (dotted lines define the standard deviation of
measurement). The bottom dash-dot line shows the size of the macropixel used
in the measurements.}
\end{figure}
%%%%%%%%%
The character of the observed speckle patterns remains stable in
the range between 30~mW and 50~mW of the pump power $ P_{p} $, as
shown in Fig.~3. The width $\Delta\Gamma$ of intensity XC function
both in the radial [Fig.~3(a)] and azimuthal [Fig.~3(b)]
directions is systematically larger than the corresponding width
of the intensity AC function. This occurs probably due to the geometry
of our experiment which is not perfectly far-field as we do not use any imaging to simplify the
setup. The drop in AC function widths
$\Delta\Gamma$
occurring for low pump powers $ P_{p} $ (below 30~mW) reflects
insufficient coherence inside the field observed for not
sufficiently intense stimulated process.
The width of the XC function $\Gamma_{is}$ remains large in this
region, as the coherence between the signal and idler fields (XC)
originates in spontaneously generated photon pairs whereas the
internal coherence (AC) of both emitted beams arises from
stimulated emission in the nonlinear process  \cite{Zou1991}.
Below 20~mW, the AC correlation-function widths are already at the
level of a single macropixel of the detector and bear no other
physical meaning than the pixelization of the detector. The dashed
line in Fig.~3 is drawn at the value of XC function width
$\Delta\Gamma_{is}$ measured at the single-photon level ($ P_{p} =
20~\mu$W). We can see that its value coincides well with the
widths $\Delta\Gamma_{is}$ of the correlation functions obtained
by classical-intensity measurements. This gives an experimental
evidence that the correlation functions obtained in single-photon
and classical-intensity measurements represent the same physical
phenomenon of PDC though observed at different physical
conditions. Whereas the twin beams at single-photon level are
described by highly nonclassical multi-mode Fock states with
completely random optical phases, the intense twin beams are well
described by multi-mode classical coherent states whose well
defined phase properties inside individual modes manifest
themselves with the occurrence of speckles\cite{Brambilla2004}.

%%%%%%%%%
\begin{figure}[tbh]  % 4
\centering{\includegraphics[width=7cm]{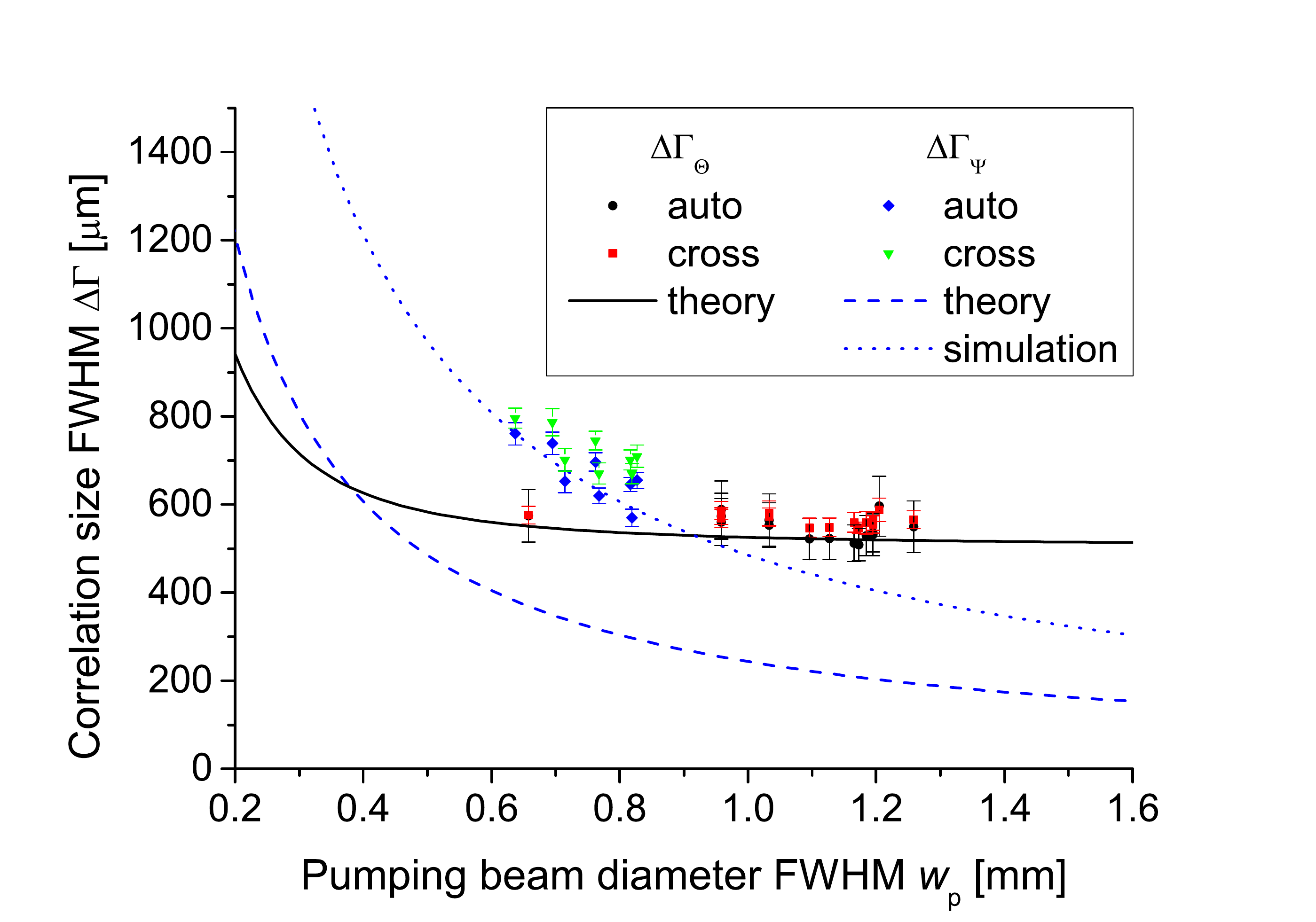}} \caption{Width
$\Delta\Gamma_{ii}$ of intensity AC functions (circles, triangles)
and $\Delta\Gamma_{is}$ of XC functions (rectangles, diamonds,
FWHM) as they depend on pump-beam diameter $ w_p $. Radial
(circles, rectangles) and azimuthal (triangles, diamonds)
dependencies are shown simultaneously. Azimuthal widths
$\Delta\Gamma_{\Psi}$ are plotted against vertical pump beam
diameter $ w_p $ while radial widths $\Delta\Gamma_{\Theta}$ are
drawn against horizontal pump-beam diameter $ w_p $. The placement
of the azimuthal points at lower values of $w_p$ stem from
ellipticity of the pump beam. Solid (radial) and dashed
(azimuthal) curves are based on a theoretical model
\cite{PerinaJr2011}. Dotted line shows a theoretical curve for a
pumping beam with spatial spectrum twice wider than an ideal
Gaussian beam (see the text).}
\end{figure}
%%%%%%%%%

Close similarity in the behavior of intensity correlation
functions in both regimes has been experimentally confirmed by
observing the behavior of correlation size as a function of the
pump-beam diameter $ w_p $. As shown in \cite{Hamar2010}, the
intensity XC function is sensitive to the pump-beam diameter $ w_p
$ inside the nonlinear crystal, especially in the azimuthal
direction. For this reason, we have investigated this dependence
in the classical-intensity regime (see Fig.~4). In agreement with
our previous results obtained at the single-photon level
\cite{Hamar2010}, the radial widths $ \Delta\Gamma_{\Theta} $ of
both intensity AC and XC functions do not significantly depend on
the pump-beam diameter $ w_p $. This is caused by the fact that
the phase-matching condition along the propagation $ z $ axis is
much more restrictive for the used BBO crystal than the
phase-matching condition in the transverse plane related to the
radial direction that encompasses the pump-beam spatial
spectrum\cite{Hamar2010} (see Fig.~4). On the other hand, a
roughly hyperbolic dependence of the widths $ \Delta\Gamma_{\Psi}
$  of intensity correlation functions on pump-beam diameter $ w_p
$ has been observed in the azimuthal direction [see Fig.~4]. This
dependence arises from the phase-matching condition along the
azimuthal direction, which crucially depends on the corresponding
pump-beam spatial spectrum. The wider the pump-beam spatial
spectrum the less strict the phase-matching conditions and so the
wider the observed widths. Also theoretical predictions for the
widths $ \Delta\Gamma $ of intensity correlation functions are
drawn in Fig.~4. They were obtained by the model that decomposes
the interacting fields into plane waves, uses Gaussian pump
temporal and spatial spectral profiles and considers
phase-matching conditions for the wave vectors along all three
directions (for more details, see \cite{PerinaJr2011}). As it
follows from Fig.~4, both experimental data and theoretical curves
are in qualitative agreement. However, a complicated multi-mode
spatial structure of the intense pump beam that was difficult to
monitor and control in the experiment did not allow us to model
the nonlinear process with the sufficient precision needed for
finding a better agreement. The complicated pump-beam structure in
its transverse plane results in a spatial spectrum wider and
richer than the Gaussian spectrum considered in the model. As a
consequence, the experimental widths of intensity correlation
functions in azimuthal direction are wider (see dotted line in
Fig.~4 for an example of the effect of a wider spatial spectrum).

\section{Conclusion}
We have shown that intensity correlation functions of twin beams
measured at single-photon level correspond to those of twin beams
with speckle patterns reached at high intensities. The
single-photon and high-intensity XC functions are equal as the
mutual coherence of signal and idler fields arises from the
generation of photon pairs. On the contrary, internal coherence of
both fields originates from stimulated emission and so AC
functions drop down to zero for weak fields. The obtained
experimental dependencies of intensity AC and XC functions of
intense speckle fields on pump-beam parameters and their
comparison with the corresponding single-photon counterparts have
confirmed this correspondence.

\section*{Acknowledgments}
Support by projects P205/12/0382 of GA \v{C}R and projects CZ.1.05/2.1.00/03.0058
and CZ.1.07/2.3.00/20.0058 of M\v{S}MT \v{C}R are acknowledged. RM thanks to IGA
PrF\_2014\_005. The support of MIUR (FIRB LiCHIS - RBFR10YQ3H) is also
acknowledged.

\end{document}